# "It was Colonel Mustard in the Study with the Candlestick": Using Artifacts to Create An Alternate Reality Game—The Unworkshop

**Alina Striner**
University of Maryland
College Park, MD 20740, USA
algol001@umd.edu

**Lennart E. Nacke**
University of Waterloo
Waterloo, Ontario, Canada
lennart.nacke@acm.org

**Elizabeth Bonsignore**
University of Maryland
College Park, MD 20740, USA
ebonsign@umd.edu

**Matthew Louis Mauriello**
University of Maryland
College Park, MD 20740, USA
mattm@cs.umd.edu

**Zachary O. Toups**
New Mexico State University
Las Cruces, NM, USA 88003
z@cs.nmsu.edu

**Carlea Holl-Jensen**
University of Maryland
College Park, MD 20740, USA
cholljen@umd.edu

**Heather Kelley**
Carnegie Mellon University
Pittsburgh, PA 15213
hkelley@andrew.cmu.edu

## Abstract
Workshops are used for academic social networking, but connections can be superficial and result in few enduring collaborations. This *unworkshop* offers a novel interactive format to create deep connections, peer-learning, and produces a technology-enhanced experience. Participants will generate interactive technological artifacts before the unworkshop, which will be used together and orchestrated at the unworkshop to engage all participants in an alternate reality game set in local places at the conference.

## ACM Classification Keywords
Games/Play; Interaction Design; Prototyping; Embodied Interaction; Storytelling

## Introduction
Within the scope of the Special Interest Group of Computer-Human Interaction (SIGCHI), conference workshops are known as gathering places for conference attendees with shared interests to meet for focused discussions. However, everyone has attended work-shops that are organized like mini-conferences, which lack sufficient discussion and networking opportunities. This *unworkshop* offers something new and exciting for the SIGCHI community that does not follow the traditional workshop format, but is focused on novel performative activities around technology development that promote networking, interaction, and creative expression via an evening murder mystery.

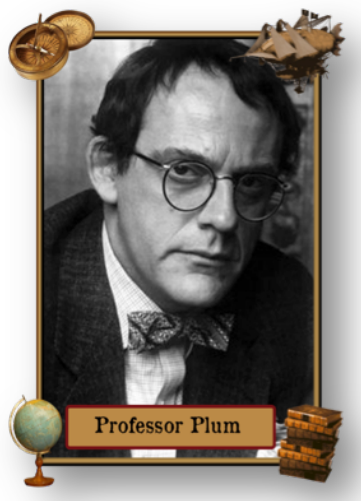

Figure 1: This sample "suspect" card, revised from the mystery game, Clue, gives a sense of one of the simplest type of character-based artifact that participants might bring. Text and/or photos that serve to extend the character's back-story could be included in this low-tech artifact.

*Sample Scenario*
*On a quiet spring evening, an unidentified man staggers into a hotel lobby in Larimer Square, Denver, and falls down dead. Who is this man? What was he doing in the hotel, and what is the mysterious technical device he had clutched in his hand? Who killed him, and why? A blinking e-textile ribbon in his lapel leads to a cafe across town, where—rumor has it—double agents congregate and secret deals are made. Anyone in the cafe might be an ally of the murdered man, or they might be enemy spies—everyone is a suspect! Further investigation only serves to raise more questions about the dead man, whose murder seems to involve secret drone squadrons located in an obscure foreign principality, covert tensions between nations, and an escalating technological conflict unfolding right here in an ordinary American city.*

*Activities and Format*
What is special about this murder mystery format [11] and why are we incorporating it into a SIGCHI Workshop? The participants of this workshop will take an interactive narrative frame like the scenario presented above and explore ways to integrate physical computing artifacts into it, with the goal of prototyping a mixed-reality experience [1] that they perform together with public spectators.

Participants will have the opportunity to discuss design issues of interest with game designers and ubiquitous computing researchers, such as how to effectively orchestrate game mechanics and narrative together into a playful experience [2]. Makers and artists who are interested in physical computing and the Internet of Things will tinker with interactive props; experimenting like *bricoleurs* [19] (i.e., people like the TV persona *MacGyver* that create solutions for a problem out of immediately available found objects) to integrate interactive, computational artifacts into the scenario [18].

The final workshop event will be a dynamic murder mystery enacted between designers and spectators at CHI, facilitating discussions about design issues and design theory associated with live gaming and interactive play [8][16]. The underlying workshop theme is to consider what design issues and questions arise from integrating physical, material computing assets (e.g., e-textiles, wearables) into an interactive narrative that is played by both designers and spectators. Most importantly, however, our workshop will be conducted as a novel collaborative design experiment whose goal is to enable a rich peer-learning opportunity and produce an engaging and enjoyable performative experience. It will offer participants a research opportunity for considering how such novel design formats might lead to stronger, sustained community ties.

## Background
Team-building exercises have often been used to establish connections and trust in academic and work environments [4]. Successful exercises create trust by asking participants to role-play and perform collaboratively to successfully complete a mission [6] or fulfill a narrative [12]. Similarly, the unconference format [1] has been used successfully to support interdisciplinary collaboration and pro-social community efforts for decades, by researchers and practitioners in fields as diverse as software engineering, library science, digital humanities, and science and technology studies [14][21].

Games facilitate natural team building and promote the development of leadership skills through goal setting, interpersonal relations, and problem solving [4]. Alt-hough collaborative games have been used to facilitate teamwork [12], improvisatory theatrical games have not only been shown to increase cooperation and build trust, but have also been used in ideation and participatory design [13].

The blurred line between playing and making has the potential to double the impact of the event: a well-designed environment can allow participants to form meaningful relationships as well as produce a memorable artifact or experience. Unlike superficial connections, professional trust is based on performance and competency [7]; thus the process of co-creation may help to facilitate deeper connections and relationships [15]. Co-creation with a new group may have even deeper impact: researchers and designers from diverse backgrounds have a variety of insights and skills that they can share during co-design, thus promoting a collaborative learning environment.

## Novelty and Impact

The CHI researchers, designers, and enthusiasts are armed with talents and insights to design meaningful user experiences. As an alternative to traditional workshops, our workshop's goal is to design a unique novel opportunity for participants to network, co-create, and learn by designing, performing, and playing an alternate reality interactive theater experience.

Rather than asking for position papers, the workshop will have a "call for mavericks," asking users to submit a proposal for a tangible or wearable artifact, accompanied by a short description of functionality and motivation. The goal of the artifact is to represent an element of a game that can be incorporated into the overarching theatrical experience that will be co-designed by participants during the workshop. Examples may include:

- *A character representation or accessory*, such as a collection of "suspect cards" that provide character back-stories (Figure 1), or an unassuming bow-tie that squirts poison (Figure 2).

- *A game mechanic that establishes rules or constraints, whether for the interactive narrative and characters, or the audience.* Examples might include a special story timer that constrains the timeframes within which characters can act or the timeframes within which spectators can give clues to some characters and hide them from others.
- *A chance mechanism* that enables opportunities for randomness, such as, an Arduino-driven "Magic-8 Ball" whose options require some type of audience participation or character action (Figure 3).
- *A way to express outcome or reward*. Examples might include an interactive device that specifies audience responses such as applause or heckling.
- *Elements that represent the game space or story world,* such as an LED constellation of key locations (Figure 4), an audio file of distinct sound effects, or a dinner glass that vibrates whenever someone walks by. As an example from an existing interactive performance, the play *Sleep No More* created an interactive Ouija board to let participants communicate with each other [20].

In addition to producing artifacts as part of their proposals, participants will be asked to provide professional and personal background that will be used to develop the characters that will inhabit the overarching workshop experience. Participants will use the artifacts they build to co-create and facilitate a hi-tech retelling of mystery-based deduction and parlour games like *Clue* [11, G2], *Mafia* [4] *Inkognito* [G1], and *Werewolf* [17] with an interactive audience, whose goal it will be to determine a killer. The design portion of the workshop will take place throughout the course of two days, with a culminating performance that will be enacted over dinner on the evening of the second day and will be open to all CHI attendees and the general public – akin to a traditional murder mystery dinner theater. A theater format has not previously been tackled in SIGCHI workshops, so research questions are: *How much structure must be imposed to make this method effective? Will it create stronger professional connections*

**Potential Character-based, Narrative-Driven Artifact**

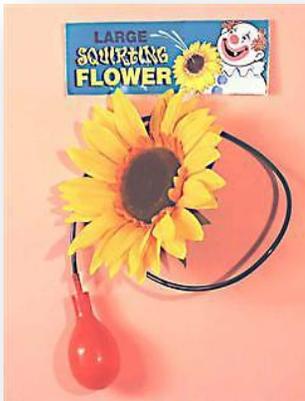

Figure 2: A flower that shoots poison is another example of a low-tech narrative-driven artifact that participants might propose to incorporate into the workshop's final interactive performance

**Potential "Chance" Mechanism Artifact**

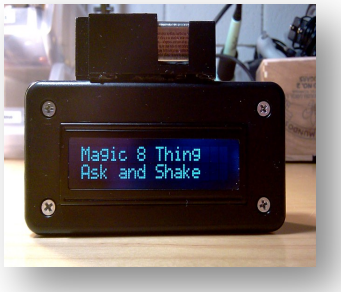

Figure 3: An arduino-style magic 8 ball, an example of a game mechanic. "Magic 8 Thing" © Mills, Pete. PeteMills.blogspot.com

than traditional workshops? Will it generate peer learning? What experience will it provide for audiences and/or spectators? How might spectator participation be influenced or expanded?

The goal of this workshop is not only to create a novel experience for participants, but also to broadly consider how outward-facing co-creation may help future research communities create more meaningful ties and develop more productive, sustainable collaborations. Consequently, a research goal for the workshop is to present the experience as an experimental field study.

## Organizers

*Alina Striner* is a third year doctoral candidate at the University of Maryland's Human Computer Interaction Lab (HCIL). Her research considers how multisensory interactivity may be used to train assessment and create engagement and immersion in theater and virtual reality environments.

*Lennart E. Nacke, Ph.D.,* is an Associate Professor for Human-Computer Interaction and Game Design at the Department of Drama and Speech Communication, affiliated with The Games Institute and Stratford Campus at the University of Waterloo. He is known for his ground-breaking gamification research, forming a new HCI research subfield. He has expanded his research program with novel visualization tools for games user research, player personality models, and health game research and development. Dr. Nacke has co-organized many workshops for CHI over the past five years (one of them the highest-cited workshop "Gamification: Using Game Design Elements in Non-Gaming Contexts"[1]); he also chaired the CHI PLAY 2014 and Gamification 2013 conferences, served as technical program co-chair for CHI PLAY 2015 and CHI Games and Play subcommittee co-chair for CHI 2017, and is currently the chair of the CHI PLAY steering committee.

---
[1] 712 citations as of writing, highest cited CHI publication in the last 5 years according to Google Scholar.

*Elizabeth Bonsignore, Ph.D.,* is a postdoctoral researcher at the University of Maryland. Her research focuses on the design of technology-mediated social experiences that promote new media literacies, arts-integrated science learning, and participatory cultures for youth. She is particularly interested in the role that multimodal narratives play in helping under-represented youth engage in life-long learning practices.

*Matthew Louis Mauriello* is a fifth-year doctoral student at the University of Maryland's Department of Computer Science. His research focuses on sustainability, educational technology, and games. He is particularly interested in how games, and elements of gamification, can encourage people to engage with important social issues.

*Zachary O. Toups, Ph.D.,* is an Assistant Professor with a focus in human-computer interaction and game design; he is director of the Play and Interactive Experiences for Learning (PIxL) Lab. His research covers various intersections of game design, mixed reality, wearable computing, and disaster response with a focus on how players collaborate and coordinate in games.

*Carlea Holl-Jensen* is a writer who's fiction has appeared in *Fairy Tale Review and Queers Destroy Fantasy!,* among others. She holds an MFA in Fiction from the University of Maryland, where she now works in the Human-Computer Interaction Lab.

*Heather Kelley* is Assistant Teaching Professor in the Entertainment Technology Center at Carnegie Mellon University and an award-winning game designer, media artist, and curator. Ms. Kelley's extensive career in game development has included design and production of virtual reality puzzle games, touchscreen vibrator controllers, AAA next-gen console games, interactive smart toys, mobile and handheld games, research games, playful museum installations, and web communities for girls.

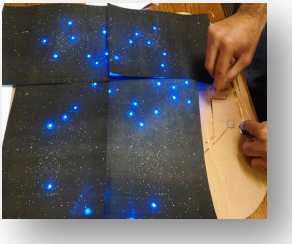

Figure 4: A LED constellation map, an example of a potential story world artifact. "StarryNight" © Malu, Meethu & Maidasani, Hitesh. cmsc838f-s14.wikispaces.com/StarryNight

**Recruiting & Pre-Workshop Plans**

CHI is an ideal venue from which to recruit participants for this uniquely interactive, improvisational theatre design format, because it is a uniquely interdisciplinary community of designers, researchers, practitioners, and students across academia and industry. We anticipate that many interested participants will come from game design and player experience sub-communities; however, we hope to attract participants from many fields beyond UX/HCI, such as ubiquitous computing, digital humanities, performance art, and film studies. We see our workshop not only as a forum on collaborative design and peer learning for the participants who will be co-designing the interactive mystery, but also as an opportunity to explore the dynamic designer-spectator relationships generated in our workshop finale, the public-play experience. Most importantly, the workshop can serve as a small field study on the challenges, opportunities, and long-term community ties that arise from this transformative workshop format.

We plan to recruit via numerous social media channels, including (but not limited to) community threads and listservs such as CHI-Announcements, Ubiquitous Computing (UbiComp), Pervasive Computing (PerCom), and Games4Change. We will take advantage of formal professional networks such as LinkedIn and the International Game Developers Association (IGDA), whether at the local level (*e.g.,* IGDA Colorado Chapter) or national level (e.g., SIGs like IGDA Game Writers or Game Education). We will invite participation from more informal practitioner networks as well, such as Hackerspace communities (*e.g.,* through https://wiki.hackerspaces.org/Communication). We will also directly invite those who have published in the area of mixed reality/pervasive game design, wearable computing, tangible computing, interactive fiction, and performance arts. Of note, we have received initial interest from several game designers and player experience researchers who are supportive of this workshop.

*Pre-Workshop Plans*
A "call for mavericks" will be issued per the details above. Interested participants will submit an artifact proof of concept (a photograph or sketch) and description of its interaction and fit into the broader narrative arc of the final murder mystery dinner theater performance. Workshop participants will be chosen based on their artifact's functionality, originality and fit. Workshop participants who submit to the Early Acceptance Round (i.e., by 21 December 2016) will receive additional feedback and direction from workshop coordinators on the potential narrative or game interaction tie-ins for their prototypes, and will thus have a better chance of being admitted.

Two weeks before the workshop takes place, accepted individuals will be invited to participate in a Google Hangout session to meet each other and to demo or describe their artifact prototypes. Participants will have a chance to read the artifact descriptions and watch prototype demo videos beforehand; however, it is important that participants are aware of one another's work before the workshop begins.

**Workshop Structure**
The workshop will be divided into two days: during the first day, participants will engage in improvisation, game icebreakers, and introductions, followed by guided ideation on the design of the game narrative and gameplay mechanics, using participant prototypes. During the second day, participants will synthesize the artifacts into the final story and gameplay elements as well as practice guiding the performance.

*Day 1*
The first day will begin with one-on-one and group improvisation icebreakers that allow participants to introduce one another and demonstrate the possible functionalities of their artifact. After the icebreakers, the group will gather together and engage in the day's primary tasks:

- Shape the murder mystery narrative using relevant features of each participant-contributed artifact.
- Finalize plans for audience roles, goals, and interactions during the performance.

During this time, the entire group will decide on key plot points, game rules, and mechanics.

After the whole-group discussion, participants will be divided into three design teams that align with a traditional dramatic arc. The Exposition Group will take a subset of the designed artifacts to establish the story. The Rising Action Group will further develop the story, with another subset of artifacts, and drive the experience towards its climax. Finally, the Resolution Group will make use of a subset of artifacts to conclude the story. Similarly, participants will use the artifacts' interactions to advance the storyline.

*Day 2 (Workshop Time Slot)*
During the second day, each group will overview how they are using artifacts to further the story and game, and then playtest each section—the presenting group will facilitate the game as story characters, and the non-presenting group will playtest as the audience to provide feedback. During second part of the day, participants will playtest the full murder mystery show, and address timing details.

*Day 2 (Performance Time)*
Our goal for the workshop's public performance is to enact it in a local restaurant venue during a cocktail hour or dinner. As the show is a form of collaboratively developed improv with game rules, we expect participants to play throughout the game, which will drive emergent outcomes. CHI attendees will be invited to the dinner, and the murder mystery will unfold around them. Rather than merely being spectators, audience members will have an opportunity to interact with objects and determine the final outcome.

### Evaluation and Post-Workshop Plans
In addition to experimenting with ways to assemble and integrate diverse digital interfaces into a hybrid play experience, a major goal of the workshop is to play-test a novel, interactive format and to consider how it promotes collaboration and creativity. We thus plan to treat the workshop as a formative field study to see how the experience affects participant collaborations.

At the start of the workshop, we will gather data via surveys, and build a social network graph [10][12] of the participants as a baseline. During the workshop, overall behavior trends will be noted and recorded subtly, so as not to get in the way of the workshop itself. Post-workshop, we will survey participants and audience members on their experience. The organizers will survey the participants again after six months. We will examine how the tie strength between participants changes, as well as the number of collaborations that were created as a result of the experience.

Once the data is collected, the organizers and any interested workshop participants will reconvene to write up the resulting data for potential publication in archived proceedings, as well as informally, on the workshop website (http://unworkshop.hcigames.com/).

### Proposed Call for Mavericks
Workshops are often used for academic social networking; however, superficial connections may result in little collaboration [8]. This workshop is our attempt at creating a new format that builds deep relationships through co-creation and peer learning. As game researchers, designers, and enthusiasts, the CHI2017 audience has unique insight into how to create meaningful games. An alternative to traditional workshops, the goal of this workshop is to design a unique opportunity for participants to network, co-create, and learn by creating a mixed reality interactive theater experience.

This two-day unworkshop combines tangible wearable interfaces with game design and improvisatory theater. Over the course of the workshop, participants will co-create a high-tech, interactive theater experience that will be played with and by an audience as a murder mystery.

Artifact submissions can fall into a number of categories, including: *Tangibles/wearables, sculpture (3D printed or otherwise), music/sound effects, and photographs/journals.* Artifacts should be made by participants, easily transportable and durable. Workshop participants will be chosen based on their artifact's functionality, originality, and fit. Submissions should be emailed to <u>unworkshop@hcigames.com.</u>